\documentclass{article}
\usepackage{preamble}

\title{Proof of Concept for Mammography Classification with Enhanced Compactness and Separability Modules}
\author{Fariza Dahes}
\date{December 2025}

\begin{document}

\maketitle
\begin{center}
    \large Extending Xia et al.'s framework beyond ConvNeXt-Tiny under non-CPU-only constraints
\end{center}

\begin{abstract}
This study presents a validation and extension of the methodological framework proposed by Xia et al. \cite{xia2025convnexttiny} for medical image classification. While their improved ConvNeXt Tiny architecture, integrating Global Average and Max Pooling fusion (GAGM), lightweight channel attention (SEVector), and Feature Smoothing Loss (FSL), demonstrated promising results on Alzheimer’s MRI under CPU friendly conditions, our work investigates its transposability to mammography classification. Using the Kaggle dataset curated by Venegas \cite{venegas2023kaggle}, which consolidates INbreast, MIAS, and DDSM mammography collections, we compare a baseline CNN, ConvNeXt Tiny, and InceptionV3 backbones enriched with GAGM and SEVector modules. Results confirm the effectiveness of GAGM and SEVector in enhancing feature discriminability and reducing false negatives, particularly for malignant cases. In our experiments, however, the Feature Smoothing Loss did not yield measurable improvements under mammography classification conditions, suggesting that its effectiveness may depend on specific architectural and computational assumptions. While Xia et al. demonstrated its value in CPU friendly settings, our results highlight its limited relevance in our context. Beyond validation, our contribution extends the original framework through multi metric evaluation (macro F1, per class recall variance, ROC/AUC), feature interpretability analysis (Grad CAM), and the development of an interactive dashboard for clinical exploration. As a perspective, we highlight the need to explore alternative approaches to improve intra class compactness and inter class separability, with the specific goal of enhancing the distinction between malignant and benign cases in mammography classification.
\end{abstract}

\section{Introduction}

\begin{figure}[H]
  \centering
  \includegraphics[width=0.9\linewidth]{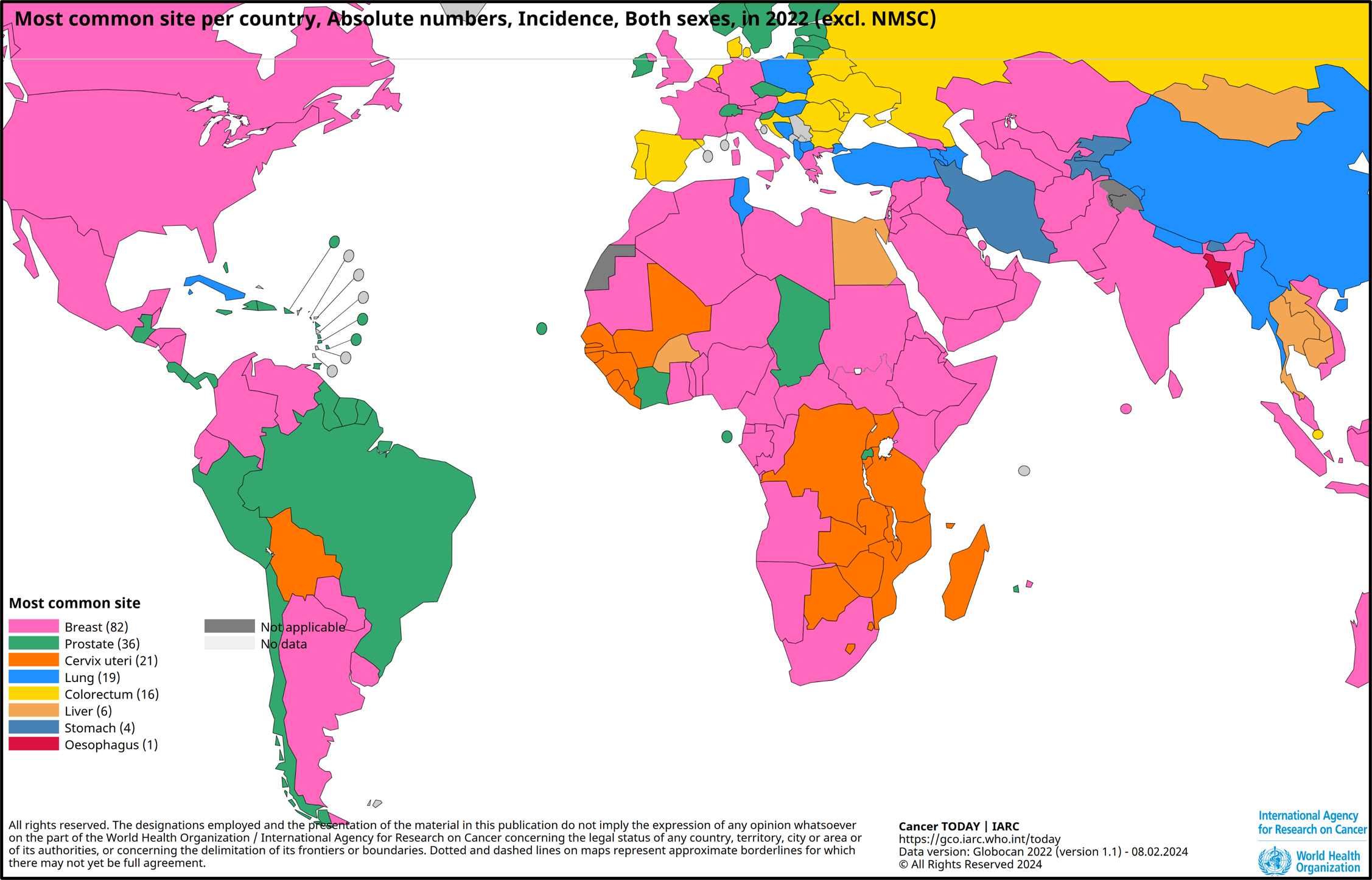}
  \caption{Most Common Cancer Site per Country (2022). Source: International Agency for Research on Cancer (IARC), WHO.}
  \label{fig:global_cancer}
\end{figure}

\begin{figure}[H]
  \centering
  \includegraphics[width=0.9\linewidth]{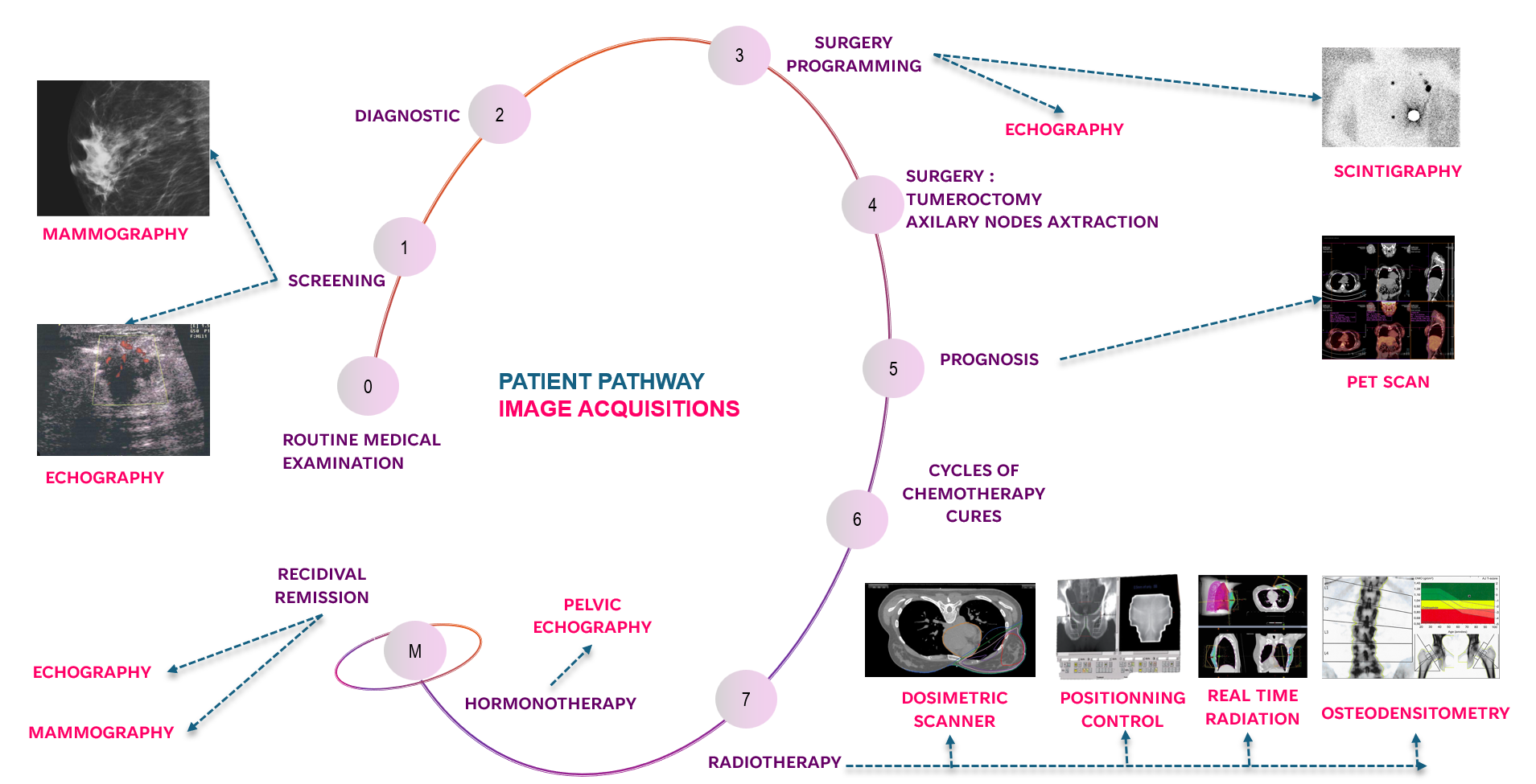}
  \caption{Patient Pathway and Imaging Modalities in Breast Cancer Care. Sequential overview of the patient journey in breast cancer diagnosis and treatment, highlighting the imaging techniques used at each stage.}
  \label{fig:patient_pathway}
\end{figure}

Breast cancer remains one of the leading causes of mortality among women worldwide (Figure~\ref{fig:global_cancer}) \cite{iarc2022gco,kim2025globalpatterns}, and early detection through mammography screening is critical for improving patient outcomes \cite{shi2025screeningvalue,henderson2024jama}. As illustrated in Figure~\ref{fig:patient_pathway}, mammography plays a central role in the initial stages of the patient care pathway—particularly during routine medical examinations and screening—before any diagnostic, surgical, or therapeutic decisions are made. This imaging modality is often the first entry point into the clinical workflow, preceding echography, scintigraphy, PET scans, and radiotherapy planning. Its strategic position in the care continuum underscores the importance of accurate and interpretable classification at this early stage.

Computer‑aided diagnosis (CAD) systems based on deep learning have shown promise in assisting radiologists by automating the classification of mammograms into normal, benign, and malignant categories \cite{hassan2022cadreview,hussain2025optimizeddl,sharma2025comparative,alantari2024cad}. However, conventional convolutional neural networks (CNNs) often face limitations in this domain, including sensitivity to class imbalance, insufficient feature separability, and reduced interpretability in clinical practice.

Recent advances in lightweight architectures have sought to address these challenges by improving efficiency and discriminative power. In particular, Xia et al. \cite{xia2025convnexttiny} proposed an improved ConvNeXt‑Tiny model tailored for medical image classification under CPU‑friendly conditions. Their framework introduced three methodological components: Global Average and Max Pooling fusion (GAGM), a lightweight channel attention mechanism (SEVector), and a Feature Smoothing Loss (FSL), achieving strong performance on Alzheimer’s MRI classification tasks.

Inspired by this work, our study investigates the transposability of these architectural innovations to mammography classification. Using the Kaggle dataset curated by Venegas \cite{venegas2023kaggle}, which consolidates INbreast, MIAS, and DDSM collections, we evaluate the effectiveness of GAGM and SEVector modules across multiple backbones (baseline CNN trained from scratch, ConvNeXt‑Tiny, and InceptionV3). Beyond validation, we extend the original framework by incorporating a multi‑metric evaluation strategy, including macro F1, per‑class recall variance, ROC/AUC analysis, and overfitting indicators. For each model, we generated classification reports, confusion matrices, learning curves, and ROC curves per class. We also performed PCA on the feature layer with the highest explained variance, and applied Grad‑CAM for interpretability \cite{selvaraju2016gradcam}, both on correctly predicted cases and critical misclassifications. This comprehensive evaluation pipeline enables a deeper understanding of model behavior and clinical relevance. In addition, we developed an interactive dashboard to support clinical exploration and facilitate model interpretability. Our objective is to assess both the strengths and limitations of these recent architectural components in a new clinical context, while highlighting perspectives for improving intra‑class compactness and inter‑class separability—key challenges in mammography classification where benign and malignant lesions often exhibit subtle visual overlap.

\section{Methods}

\begin{figure}[H]
  \centering
  \includegraphics[width=0.9\linewidth]{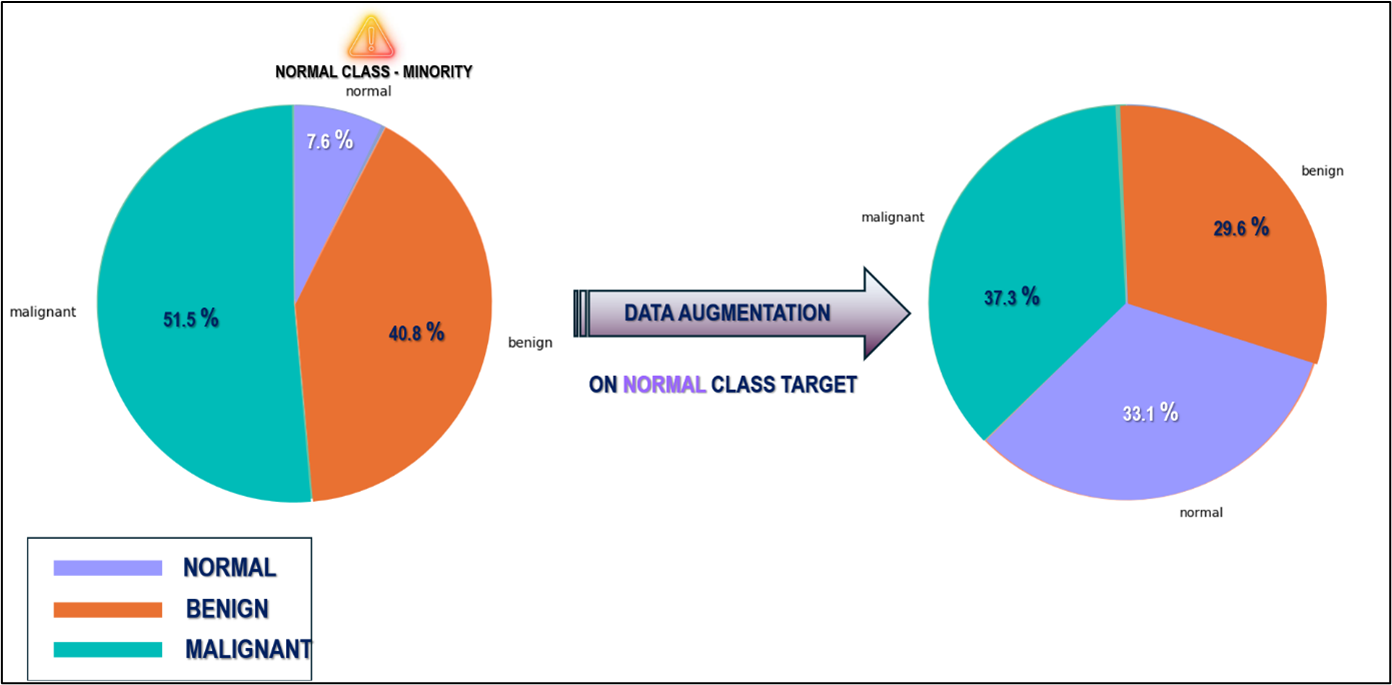}
  \caption{Class Distribution Before and After Targeted Augmentation — Comparison of class proportions before and after targeted data augmentation on the normal class. Initially underrepresented (7.6\%), the normal class was synthetically enriched to reach 33.1\%, reducing class imbalance and improving model exposure to low‑frequency patterns. This strategy supports better generalization and recall for the minority class, which is critical in screening contexts where false negatives carry high clinical risk.}
  \label{fig:data_augmentation}
\end{figure}

\textbf{Dataset and Preprocessing}  
The models developed in this proof of concept were trained on a publicly available dataset hosted on Kaggle, which consolidates mammograms from three widely recognized medical repositories: INbreast, MIAS, and DDSM \cite{venegas2023kaggle}. The dataset provides images preprocessed with CLAHE (Contrast Limited Adaptive Histogram Equalization), ensuring homogeneous quality and reducing preparation complexity. Images are categorized into three classes—normal, benign, and malignant—with a notable imbalance due to the underrepresentation of the normal class (2026 images compared to 10,900 benign and 13,700 malignant). To mitigate this bias, targeted data augmentation was applied to the normal class (Figure~\ref{fig:data_augmentation}). For demonstration purposes, 150 images were extracted to feed a blind‑test dashboard, while the remainder was split into training (80\%) and testing (20\%) subsets, followed by a stratified train/validation split to preserve class distribution.
\textbf{Architectural Framework}  
The methodological framework was inspired by Xia et al. \cite{xia2025convnexttiny}, who proposed an improved ConvNeXt‑Tiny architecture for medical image classification under CPU‑friendly conditions. Their approach introduced three modules: Global Average and Max Pooling fusion (GAGM), SEVector channel attention, and Feature Smoothing Loss (FSL).  

To ensure reproducibility, these methodological components were translated into concrete Keras/TensorFlow layers and integrated into the backbone architectures.  

\textbf{GAGM module}  
The rationale for pooling fusion lies in the complementary roles of GAP and GMP. GAP captures global statistics that stabilize intra‑class representations by reducing variance within clusters, while GMP emphasizes salient activations that enhance inter‑class separability by highlighting discriminative regions. Their fusion, formalized as $[V_{\text{avg}} ; U_{\text{max}}]$, provides a balanced representation particularly relevant in medical imaging where benign and malignant lesions often exhibit subtle overlap. As implemented in GAGM \cite{xia2025convnexttiny} or DCAF \cite{li2025resgdanet}, this strategy simultaneously improves intra‑class compactness and inter‑class separability—two critical properties for reliable clinical deployment.

\[
V_{\text{avg}} = \mathrm{GAP}(F), \quad
U_{\text{max}} = \mathrm{GMP}(F), \quad
U_{\text{fused}} = [V_{\text{avg}} ; U_{\text{max}}]
\]

where $F \in \mathbb{R}^{C \times H \times W}$ is the input feature map.  

In Keras:  
\begin{lstlisting}
# --- Dual pooling : GAP + GMP ---
gap = layers.GlobalAveragePooling2D()(x)
gmp = layers.GlobalMaxPooling2D()(x)
fused = layers.Concatenate()([gap, gmp])
\end{lstlisting}

\textbf{SEVector module}  
The SEVector module proposed by Xia et al. \cite{xia2025convnexttiny} should be regarded as a lightweight adaptation of the original Squeeze‑and‑Excitation mechanism introduced by Hu et al. \cite{hu2017senet}. While Hu et al. explicitly modeled channel interdependencies to enhance representational power, Xia et al. reformulated this principle into a simplified variant designed to reduce computational overhead under CPU‑friendly conditions, while preserving the core idea of channel‑wise recalibration.  

The Squeeze‑and‑Excitation Vector mechanism was applied to the fused vector $U_{\text{fused}}$, following the transformation:

\[
w = \sigma \big( W_{2} \cdot \mathrm{ReLU}(W_{1} \cdot U_{\text{fused}}) \big), 
\quad V_{\text{att}} = U_{\text{fused}} \otimes w
\]

where $w$ is the learned channel attention vector and $\otimes$ denotes element‑wise multiplication.  

This was implemented in Keras as:

\begin{lstlisting}
# --- SEVector ---
channels = fused.shape[-1]
compressed = max(8, channels // reduction_ratio)
se = layers.Dense(compressed, activation='relu')(fused)
se = layers.Dense(channels, activation='sigmoid')(se)
attended = layers.Multiply()([fused, se])
\end{lstlisting}

The squeeze step was implemented as \texttt{Dense(units=2C/r, activation='relu')}, the excitation step as \texttt{Dense(units=2C, activation='sigmoid')}, and the recalibration as \texttt{Multiply()}.

\textbf{Feature Smoothing Loss (FSL)}  
Xia et al. \cite{xia2025convnexttiny} introduced the Feature Smoothing Loss (FSL) to enhance intra‑class consistency. As they note, its design philosophy is similar to the Center Loss \cite{wen2016discriminative}, but without explicitly maintaining center parameters. Instead, FSL dynamically computes class centroids within each mini‑batch and penalizes deviations from these centroids, thereby improving feature discriminability under CPU‑friendly conditions.  

The FSL module aims to reduce intra‑class variance by penalizing deviations from the class mean feature vector. It is defined as:

\[
\bar{f}_{c} = \frac{1}{N_{c}} \sum_{i=1}^{N_{c}} f_{ci}, \quad
L_{fs} = \frac{1}{C} \sum_{c=1}^{C} \frac{1}{N_{c}} \sum_{i=1}^{N_{c}} \| f_{ci} - \bar{f}_{c} \|^{2}
\]

and integrated into the total loss as:

\[
L = L_{CE} + \lambda_{fs} L_{fs}
\]

In practice, this required managing multiple outputs and computing class‑wise feature statistics during training. Despite a documented implementation, the added complexity and lack of significant performance gain led us to exclude FSL from the final pipeline. GAGM and SEVector were retained as effective enhancements.

\textbf{Models and Training Strategy}  
Three models were explored: a baseline CNN trained from scratch, an improved ConvNeXt‑Tiny, and an improved InceptionV3. In line with Xia et al. \cite{xia2025convnexttiny}, who evaluated ConvNeXt‑Tiny for Alzheimer’s disease classification from MRI, we also investigate this backbone for mammography classification. This choice is further supported by other studies that have successfully applied ConvNeXt architectures to medical imaging tasks, including segmentation of CT/MRI scans \cite{roy2023mednext, hatamizadeh2022unetrpp}. Likewise, InceptionV3 has been widely adopted in medical image classification, for example in brain tumor MRI and diabetic retinopathy detection \cite{alam2024smote}, which validates its relevance for our study.  

For the latter two, ImageNet \cite{deng2014imagenet} pretraining was retained with \texttt{include top=False}, and partial fine‑tuning was applied by unfreezing the last 20 layers. Each backbone was extended with GAGM and SEVector modules, followed by a custom classification head composed of a Dense layer (256 units, ReLU), Dropout, and a final Softmax layer adapted to the three classes. Models were compiled with the Adam optimizer (learning rate $10^{-4}$), categorical crossentropy loss, and accuracy as the primary metric.  

\textbf{Regularization and Tracking}  
To ensure training stability and limit overfitting, several mechanisms were integrated: stratified validation splits, callbacks such as \texttt{ReduceLROnPlateau} (patience=5, factor=0.5) and \texttt{EarlyStopping} with best‑weights restoration, as well as Dropout \cite{srivastava2012dropout} and BatchNormalization \cite{ioffe2015batchnorm} layers. Batch size was adjusted to balance GPU memory constraints and gradient stability. All experiments were systematically tracked, with export of metrics (accuracy, loss, macro‑F1, recall min/std, overfitting indicators), checkpoints, and learning curves.  

\textbf{Evaluation and Interpretability}  
Evaluation combined global and dispersion metrics with visual diagnostics. For each model, classification reports, confusion matrices, ROC curves per class, and PCA projections of the most informative feature layer were generated to assess discriminative capacity and inter‑class separability. Interpretability was addressed through Grad‑CAM \cite{selvaraju2016gradcam}, a widely adopted method for visual explanations of convolutional neural networks, applied both to correctly predicted cases with high confidence and to critical misclassifications (e.g., malignant cases misclassified as normal or benign). This methodological pipeline ensured reproducibility, comparability, and clinical relevance, while highlighting the strengths and limitations of each backbone in mammography classification.

\section{Results}

\subsection{Definition of Metrics Used}
To evaluate the three models—Improved InceptionV3, Improved ConvNeXt‑Tiny (ICNT), and a custom baseline CNN—we relied on a comprehensive set of metrics designed to capture not only raw performance but also equity across classes and robustness against overfitting.

\begin{itemize}
    \item \textbf{Accuracy} measures the overall proportion of correct predictions across all classes. While intuitive, it may obscure imbalances when certain categories are over‑ or under‑represented.
    \item \textbf{Loss} corresponds to the value of the cost function (categorical cross‑entropy), quantifying the discrepancy between predicted and true labels. Lower loss indicates better model fit.
    \item \textbf{Macro F1} is the unweighted mean of F1 scores computed per class. Unlike weighted F1, it does not account for class imbalance, making it suitable for assessing fairness across categories.
    \item \textbf{F1 Standard Deviation (F1\_std)} captures the variability of F1 scores between classes. A low F1\_std suggests that the model performs consistently across all categories.
    \item \textbf{Minimum Recall (recall\_min)} identifies the lowest recall value among all classes. It highlights the weakest coverage and reveals whether a class is being neglected.
    \item \textbf{Recall Standard Deviation (recall\_std)} measures the dispersion of recall scores across classes. A low recall\_std indicates balanced coverage and equitable treatment.
    \item \textbf{Overfit Accuracy (overfit\_acc)} is the difference between training and test accuracy. A large gap signals overfitting and poor generalization.
    \item \textbf{Overfit F1 (overfit\_f1)} compares macro F1 scores between training and test sets, revealing whether class balance is preserved outside the training context.
    \item \textbf{Overfit Loss (overfit\_loss)} quantifies the difference in loss between training and test sets. A high value suggests excessive adaptation to training data and limited generalization.
\end{itemize}

This metric suite enables a nuanced evaluation of model behavior, going beyond raw accuracy to assess robustness, equity, and generalization capacity—critical aspects in clinical classification tasks such as mammography.
\subsection{Global Metrics Comparison}
To compare the three models—Improved InceptionV3, Improved ConvNeXt‑Tiny (ICNT), and the baseline CNN—we first examined their performance using pairwise metrics, then extended the analysis to multidimensional triplets through radar plots. This dual perspective highlights both raw performance and the trade‑offs between equity, coverage, and robustness.  

To support this comparative analysis, we first present a consolidated table of global metrics for the three models (Table~\ref{table1_global_performance}), including accuracy, loss, macro F1, recall dispersion indicators, and overfitting measures. This tabular overview provides a quantitative foundation for the subsequent visual comparisons.
\begin{table}[H]
  \centering
  \includegraphics[width=\linewidth]{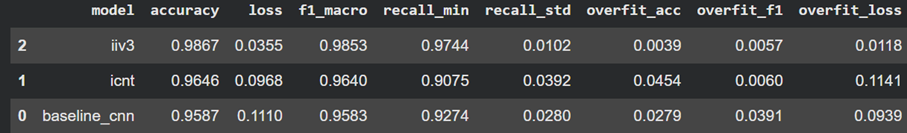}
  \caption{Global Performance and Overfitting Metrics for the Three Models — Comparison of Improved InceptionV3 (IIV3), Improved ConvNeXt‑Tiny (ICNT), and baseline CNN across key evaluation metrics. IIV3 achieves the highest accuracy and macro F1, with minimal recall dispersion and overfitting indicators. ICNT shows competitive performance but suffers from uneven class coverage and higher overfit loss. The baseline CNN offers stable but lower scores, serving as a reference point.}
  \label{table1_global_performance}
\end{table}

\begin{figure}[H]
  \centering
  \includegraphics[width=\linewidth]{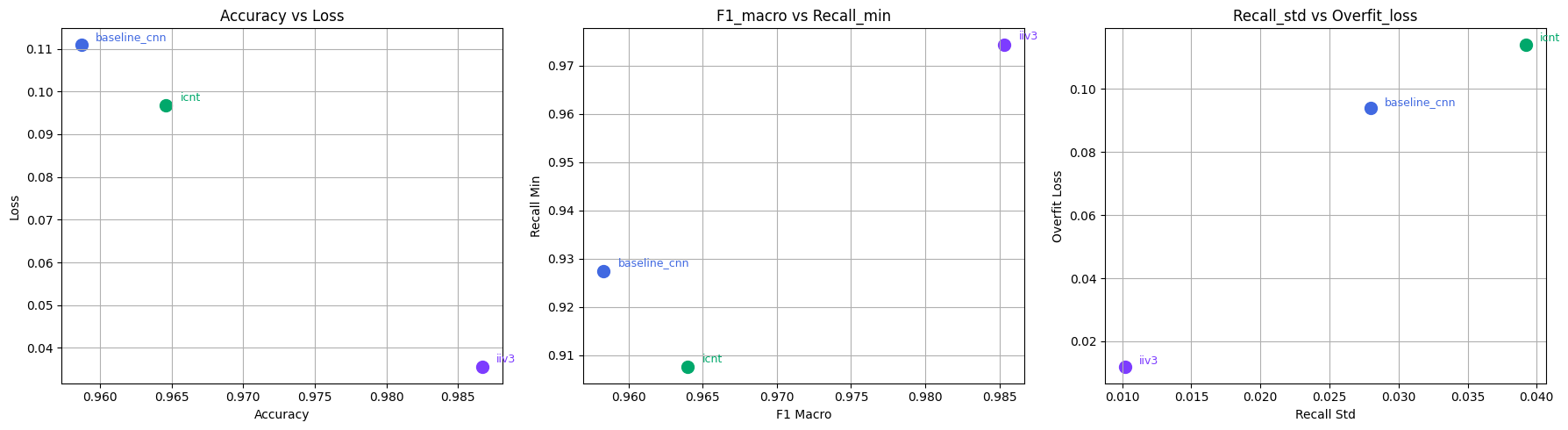}
  \caption{Pairwise Scatter Plots Comparing Model Performance — Accuracy vs Loss, Macro F1 vs Minimum Recall, and Recall Dispersion vs Overfitting Loss. These visualizations highlight the trade‑offs between precision, robustness, and equity across the three models.}
  \label{fig4_pairwise_scatter_plot}
\end{figure}

\textbf{Pairwise Metrics (Scatter Plots)}  
Scatter plots (Figure~\ref{fig4_pairwise_scatter_plot}) were used to visualize compromises between accuracy and loss, macro F1 and minimum recall, and recall dispersion versus overfitting loss. These comparisons reveal distinct profiles. ICNT achieved strong overall accuracy and macro F1, but its lower \texttt{recall\_min} and higher \texttt{recall\_std} indicate that one class was poorly covered and predictions were uneven across categories. The baseline CNN, while less performant in terms of accuracy and F1, exhibited more homogeneous behavior, with moderate dispersion and better equity between classes. InceptionV3 consistently occupied the optimal regions of the scatter plots: high accuracy with low loss, strong macro F1 combined with high \texttt{recall\_min}, and minimal recall dispersion coupled with low overfitting loss. This confirms that it is both precise and robust, without sacrificing equity.

\begin{figure}[H]
  \centering
  \includegraphics[width=\linewidth]{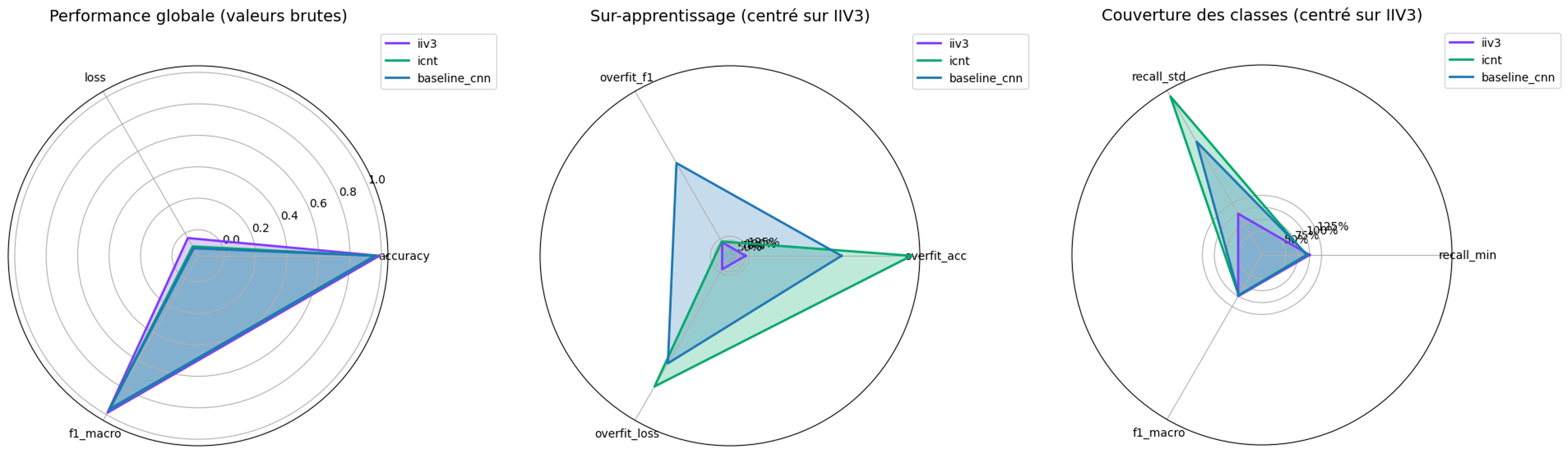}
  \caption{Radar Plots Comparing Global Performance, Class Coverage, and Overfitting Across Models. Improved InceptionV3 (IIV3) consistently forms near‑ideal polygons, reflecting high precision, equitable class treatment, and robust generalization. ICNT shows competitive global performance but suffers from recall dispersion and overfitting. The baseline CNN offers balanced but lower scores, serving as a stable reference.}
  \label{fig5_radar_plot_comparison}
\end{figure}

\textbf{Triplet Metrics (Radar Plots)}  
Radar plots (Figure~\ref{fig5_radar_plot_comparison}) provided a multidimensional view of performance, covering three complementary aspects: global performance (accuracy, loss, macro F1), class coverage (recall\_min, recall\_std, macro F1), and overfitting (differences in accuracy, F1, and loss between training and validation). InceptionV3 formed nearly perfect polygons across all triplets, excelling simultaneously in precision, equity, and generalization. ICNT remained competitive in global performance but showed instability in recall dispersion and overfitting indicators. The baseline CNN produced balanced but lower scores, serving as a stable reference point.  

\textbf{Synthesis}  
Taken together, scatter and radar analyses converge on the same conclusion: Improved InceptionV3 is the most performant, equitable, and robust model. It achieves high accuracy and F1 scores, ensures consistent coverage across classes, and generalizes well without overfitting. ICNT offers solid average performance but suffers from variability and reduced robustness, while the baseline CNN remains regular but limited. These results naturally orient the choice of model toward Improved InceptionV3 for subsequent experiments.
\subsection{Discriminative Metrics (ROC Curves)}
\begin{figure}[H]
  \centering
  \includegraphics[width=\linewidth]{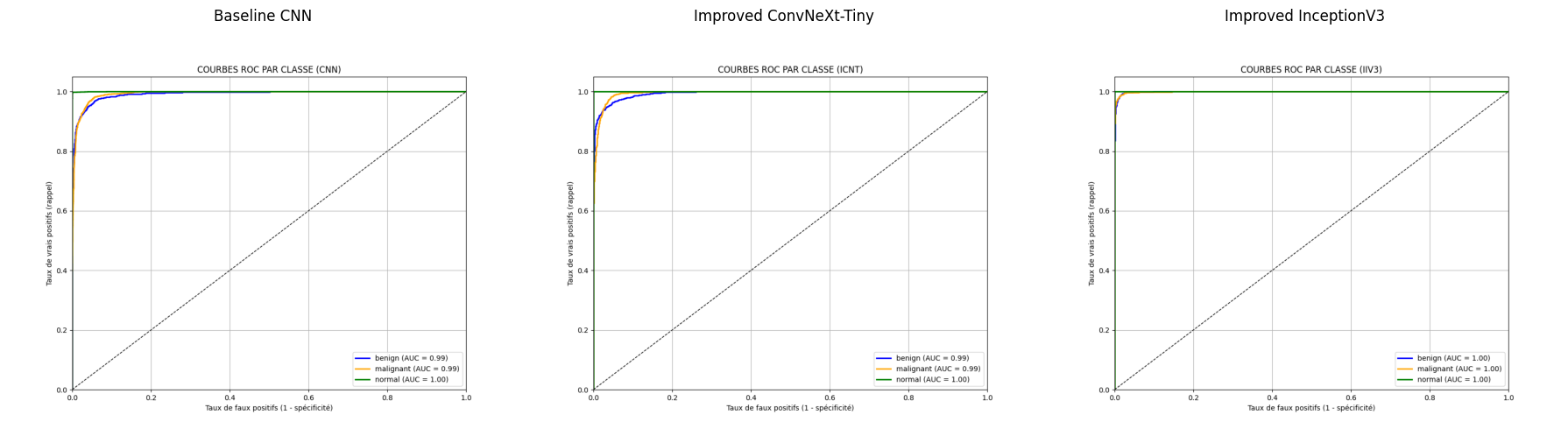}
  \caption{ROC Curves by Class for Each Model. Receiver Operating Characteristic (ROC) curves comparing the discriminative capacity of baseline CNN, Improved ConvNeXt‑Tiny (ICNT), and Improved InceptionV3 across the three diagnostic classes: normal, benign, and malignant. AUC values consistently exceed 0.95, with InceptionV3 achieving perfect scores across all categories.}
  \label{fig6_roc_curves}
\end{figure}

Beyond global performance indicators, the discriminative capacity of the models was assessed using Receiver Operating Characteristic (ROC) curves and the corresponding Area Under the Curve (AUC) values (Figure~\ref{fig6_roc_curves}). These metrics evaluate the ability of a classifier to separate classes independently of the decision threshold, which is particularly relevant in clinical contexts where false positives and false negatives carry different implications.  

\textbf{Interpretation of ROC curves.}  
The ROC curve plots the true positive rate (sensitivity) against the false positive rate (1 – specificity). An ideal model lies in the upper left corner, combining high sensitivity with low false alarm rates. AUC values close to 1.0 indicate excellent discriminative ability, while values near 0.5 correspond to random behavior.  

\textbf{Results across classes.}  
All three models demonstrated strong discriminative performance, with AUC values consistently above 0.95. For the malignant class, both the baseline CNN and ICNT achieved AUC $\approx$ 0.99, while Improved InceptionV3 reached a perfect score of 1.00, ensuring optimal cancer detection without compromising specificity. For the benign class, the same pattern was observed: baseline CNN and ICNT achieved AUC $\approx$ 0.99, while InceptionV3 again reached 1.00. For the normal class, all models achieved perfect separation (AUC = 1.00), confirming that healthy tissue was consistently recognized without false alarms.  

\textbf{Synthesis}  
These results highlight that all models are clinically reliable in distinguishing between normal, benign, and malignant cases. However, Improved InceptionV3 stands out by achieving perfect AUC across all three categories, offering the strongest guarantee of diagnostic robustness. While the differences with ICNT and the baseline CNN are numerically small, even marginal gains in malignant detection are critical in medical practice, where missing a cancer case can have severe consequences.
\subsection{Inter-Class Equity and Dispersion}
Beyond global performance, it is essential to assess whether models treat all classes with comparable effectiveness. Equity was evaluated using both macro‑averaged metrics and dispersion indicators derived from classification reports.  

\textbf{Computation of dispersion.}  
For each model, mean and standard deviation values of F1 scores and recall were calculated across classes (Table~\ref{table2_dispersion_metrics}). These statistics provide insight into the homogeneity of predictions: a low standard deviation indicates that the model performs consistently across categories, while a high value suggests uneven coverage.  

\begin{figure}[H]
  \centering
  \includegraphics[width=\linewidth]{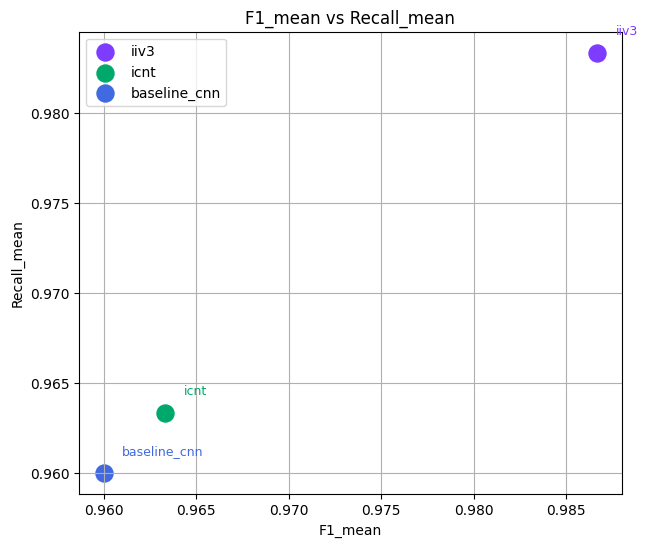}
  \caption{Inter‑Class Performance Comparison Using F1 Mean and Recall Mean. Scatter plots illustrating the relative positioning of baseline CNN, Improved ConvNeXt‑Tiny (ICNT), and Improved InceptionV3 across average F1 and recall values. InceptionV3 consistently achieves the highest averages, confirming its superiority in balanced performance.}
  \label{fig7_f1mean_vs_recallmean}
\end{figure}

\begin{table}[H]
  \centering
  \includegraphics[width=\linewidth]{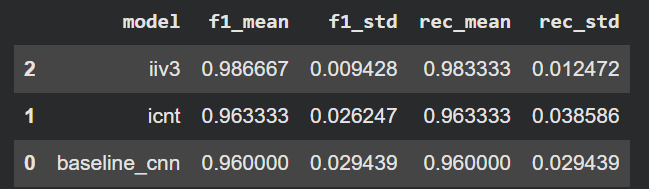}
  \caption{Inter‑Class Equity and Dispersion Metrics Across Models. Mean and standard deviation values of F1 scores and recall highlight the homogeneity of predictions. Lower dispersion indicates consistent performance across categories, while higher values reveal uneven coverage.}
  \label{table2_dispersion_metrics}
\end{table}
\textbf{Visualization and Interpretation}  
Scatter plots (Figure~\ref{fig7_f1mean_vs_recallmean}) of F1\_mean versus Recall\_mean confirmed the superiority of Improved InceptionV3, which achieved the highest averages ($>$0.98) across both metrics. ICNT outperformed the baseline CNN in mean values, but this advantage was tempered by higher variability.  

\begin{figure}[H]
  \centering
  \includegraphics[width=\linewidth]{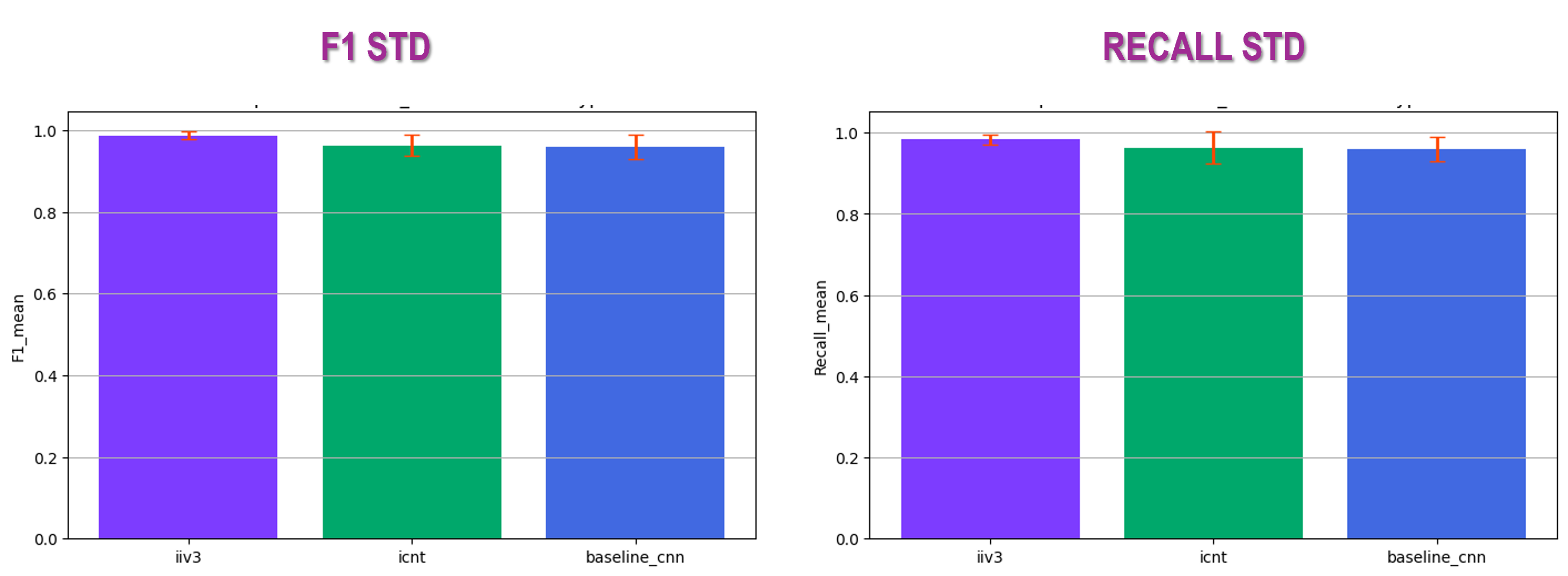}
  \caption{Bar Charts of F1 and Recall Means with Dispersion Indicators. Error bars highlight variability across classes. ICNT shows greater dispersion, particularly in recall, while the baseline CNN demonstrates lower variability despite lower overall performance. InceptionV3 combines high averages with minimal dispersion, reflecting robust and equitable predictions.}
  \label{fig8_f1std_recallstd}
\end{figure}

Barplots with error bars (Figure~\ref{fig8_f1std_recallstd}) further highlighted these differences: ICNT exhibited greater dispersion, particularly in recall, suggesting that one or more classes were less reliably captured. In contrast, the baseline CNN, although less performant overall ($\approx$0.96), showed lower variability, reflecting more homogeneous predictions.  

\begin{figure}[H]
  \centering
  \includegraphics[width=\linewidth]{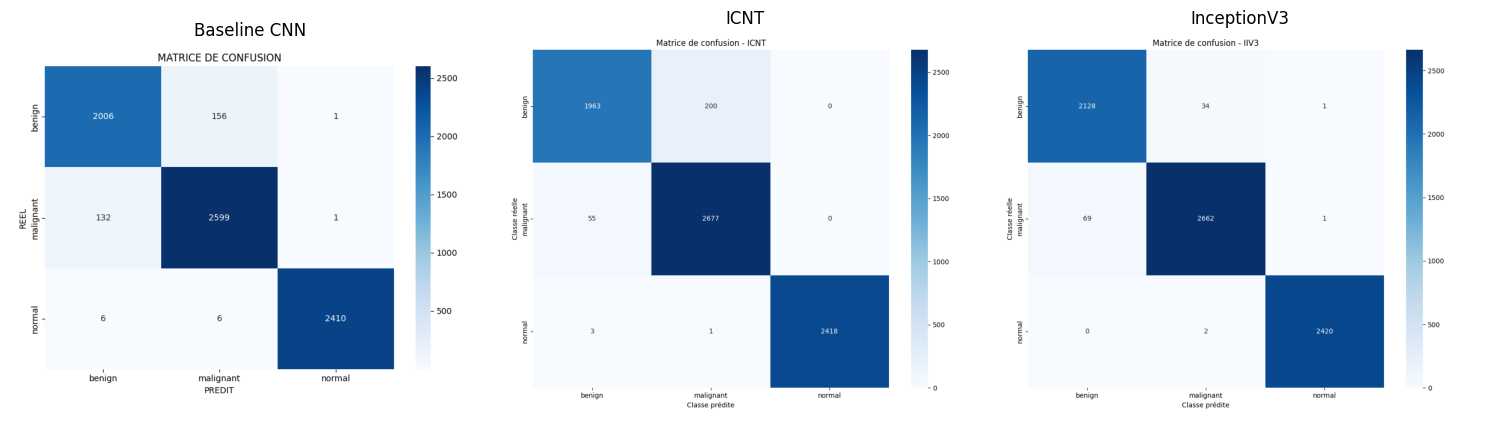}
  \caption{Confusion Matrices Showing Class‑Specific Prediction Errors Across Models. Baseline CNN reveals confusion between benign and malignant categories, ICNT improves malignant separation but sacrifices benign precision, while Improved InceptionV3 achieves near‑perfect classification with balanced precision across all classes.}
  \label{fig9_confusion_matrices}
\end{figure}

\textbf{Confusion Matrix Analysis}  
To complement these quantitative insights, confusion matrices (Figure~\ref{fig9_confusion_matrices}) were examined to visualize the distribution of errors across classes. The baseline CNN showed strong performance on the normal class (99.5\% accuracy) and acceptable results for malignant cases ($\approx$95.1\%), but revealed significant confusion between benign and malignant categories (156 and 132 misclassifications respectively). ICNT improved separation of malignant cases ($\approx$98\%) and reduced inter‑class confusion, but at the cost of lower precision on benign samples ($\approx$90.7\%). InceptionV3 achieved near‑perfect classification across all three categories, with minimal residual confusion and balanced precision: $\approx$98.4\% for benign, $\approx$97.4\% for malignant, and $\approx$99.9\% for normal.  

\textbf{Synthesis}  
Taken together, these analyses reveal that Improved InceptionV3 is not only the most accurate but also the most equitable model, combining high average scores with minimal dispersion and balanced error distribution. ICNT demonstrates stronger mean performance than the baseline CNN but suffers from instability across classes, while the baseline CNN remains more regular yet limited in competitiveness. These findings underscore the importance of considering both averages and variability when evaluating clinical models, as equity across classes is critical for reliable deployment in practice.
\subsection{Training Stability and Efficiency}
\begin{figure}[H]
  \centering
  \includegraphics[width=\linewidth]{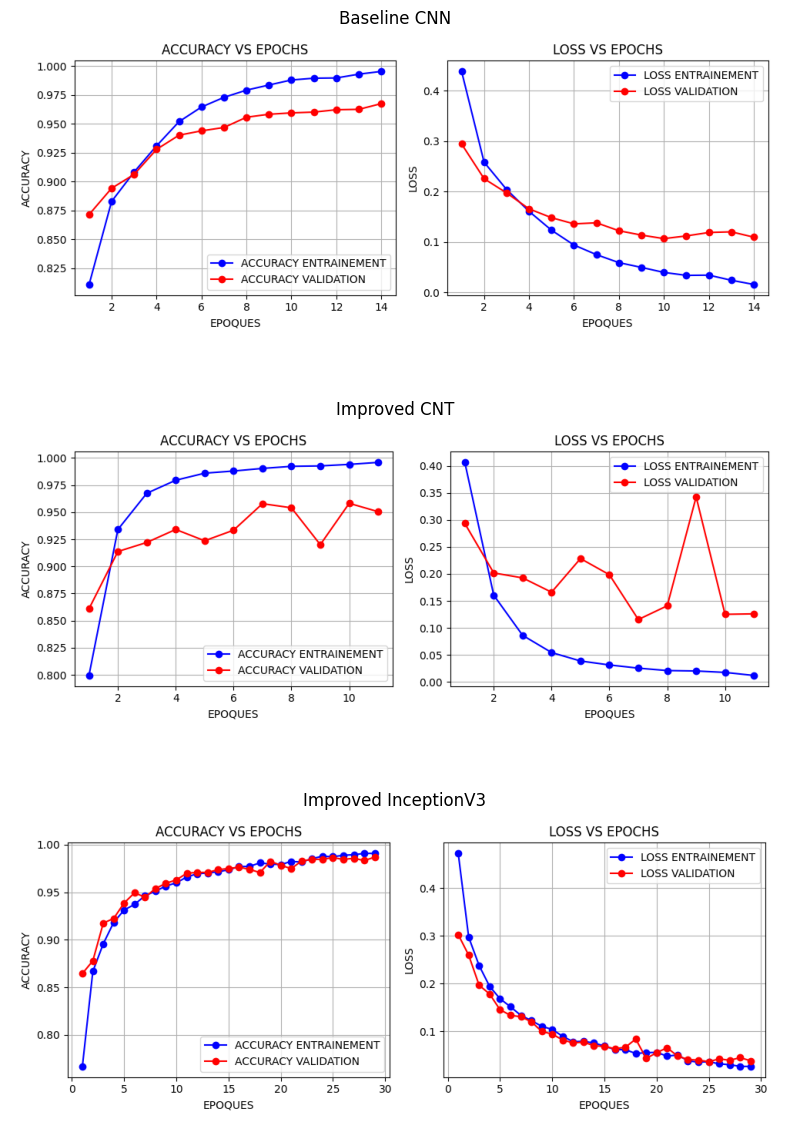}
  \caption{Learning Curves Highlighting Training Dynamics and Generalization. Accuracy and loss trajectories for baseline CNN, Improved ConvNeXt‑Tiny (ICNT), and Improved InceptionV3 across training and validation sets. InceptionV3 shows smooth convergence with minimal overfitting, ICNT reveals instability in validation metrics, while the baseline CNN remains regular but limited in optimization capacity.}
  \label{fig10_learning_curves}
\end{figure}

To assess the learning dynamics of each model, we analyzed the evolution of accuracy and loss over training epochs for both training and validation sets. These curves (Figure~\ref{fig10_learning_curves}) provide insight into convergence behavior, generalization capacity, and potential signs of overfitting.  

\textbf{Baseline CNN.}  
The baseline model exhibited steady improvement in both accuracy and loss across 14 epochs. Training and validation curves remained close, indicating regular learning behavior and limited overfitting. However, the final performance plateaued below that of the other models, suggesting limited capacity for further optimization.  

\textbf{Improved ConvNeXt Tiny (ICNT).}  
ICNT showed rapid convergence in training accuracy, reaching near‑perfect scores within 11 epochs. However, validation accuracy fluctuated, and validation loss remained unstable, revealing signs of overfitting. The gap between training and validation curves suggests that the model may have adapted too closely to the training data, compromising its generalization.  

\textbf{Improved InceptionV3.}  
InceptionV3 demonstrated the most stable and efficient learning trajectory. Over 30 epochs, both training and validation accuracy increased smoothly and converged near 1.0. Loss curves mirrored this behavior, with consistent decreases and minimal divergence between training and validation. This indicates robust convergence, strong generalization, and effective regularization.  

\textbf{Synthesis}  
Among the three models, Improved InceptionV3 exhibited the most stable and efficient training behavior, with smooth convergence and minimal overfitting. ICNT achieved high training performance but showed instability in validation, while the baseline CNN remained regular but limited in capacity. These observations reinforce the superiority of InceptionV3 not only in final metrics but also in learning dynamics.  

\textbf{Transition to Discussion}  
The comparative analysis confirmed that Improved InceptionV3 consistently outperforms the other models across all quantitative metrics, making it the natural champion in this experimental setting. Yet, model performance alone does not determine suitability for deployment. In practice, the choice of architecture depends on strategic priorities shaped by production constraints such as inference time, scalability, maintenance, and explainability.  

The scenarios outlined—initial diagnosis, longitudinal follow‑up, mass screening, and exploratory research—illustrate how different models may be favored depending on operational needs. InceptionV3 remains optimal for diagnostic accuracy, but lighter models such as the baseline CNN or ICNT may prove more practical in contexts where speed, resource efficiency, or iterative experimentation are prioritized.  

Building on this perspective, the following discussion explores two complementary axes that go beyond raw metrics: the structure and separability of latent spaces, analyzed through PCA and t‑SNE projections, and the explainability of predictions, examined with Grad‑CAM visualizations. Together, these analyses provide deeper insight into how models organize information internally and how their decisions can be interpreted in clinical practice.

\section{Discussion}

\subsection{Latent Space Structure and Separability (ACP)}
Beyond quantitative metrics, the organization of latent spaces provides critical insight into how models internally structure information. Principal Component Analysis (PCA) and t‑SNE projections were used to visualize the compactness and separability of class representations in the test set. These analyses highlight differences in how the baseline CNN, ICNT, and Improved InceptionV3 encode features and distribute classes in reduced dimensions.  

\begin{figure}[H]
  \centering
  \includegraphics[width=\linewidth]{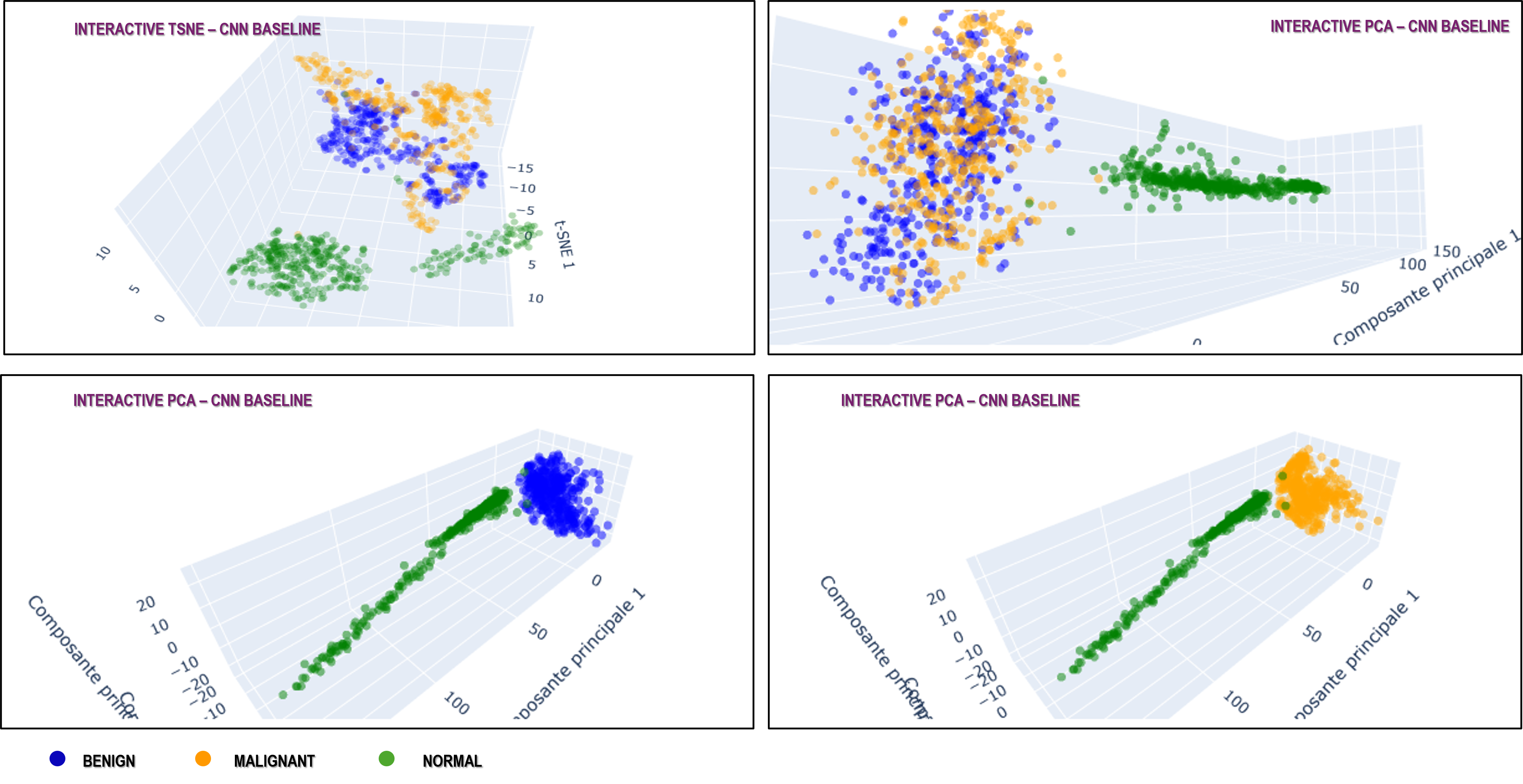}
  \caption{Latent Space Projections of CNN Baseline Using PCA and t‑SNE. The normal class forms a distinct cluster, while benign and malignant cases overlap, revealing limited discriminative capacity in the latent space.}
  \label{fig11_separability_cnn}
\end{figure}

\textbf{Baseline CNN.}  
The baseline model achieved a relatively high cumulative variance explained by the first three principal components ($\approx$87\%), suggesting that much of the feature variability can be captured in a low‑dimensional space. However, PCA and t‑SNE projections (Figure~\ref{fig11_separability_cnn}) revealed overlapping clusters, particularly between benign and malignant cases. While the normal class formed a distinct group, the separation between benign and malignant remained imperfect, reflecting limited discriminative capacity in the latent space.  

\begin{figure}[H]
  \centering
  \includegraphics[width=\linewidth]{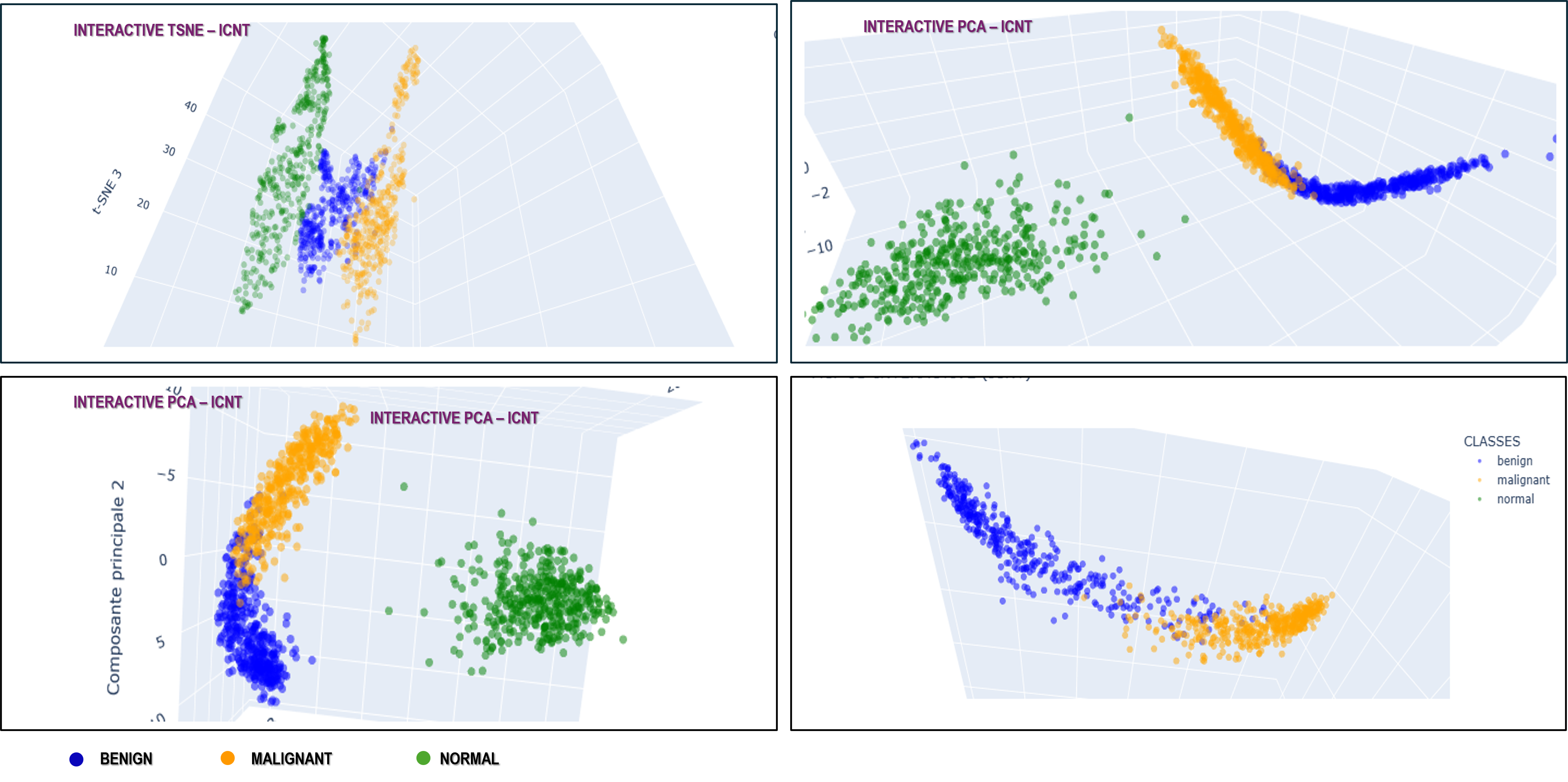}
  \caption{Latent Space Projections of ICNT Using PCA and t‑SNE. Malignant, benign, and normal clusters are more distinctly separated compared to the baseline CNN, though some dispersion remains between benign and malignant cases.}
  \label{fig12_separability_icnt}
\end{figure}

\textbf{Improved ConvNeXt Tiny (ICNT).}  
ICNT offered a more structured latent space than the baseline CNN. PCA curves (Figure~\ref{fig12_separability_icnt}) from deeper layers (e.g., \texttt{dense\_4} and \texttt{dense\_6}) showed strong variance capture, and both static and interactive 3D projections revealed clearer boundaries between classes. The t‑SNE visualization confirmed this improvement, with malignant, benign, and normal clusters appearing more distinctly separated and less overlapping than in the baseline CNN. Nevertheless, some dispersion persisted, particularly between benign and malignant, indicating that while ICNT improves separability, it does not fully eliminate ambiguity.  

\begin{figure}[H]
  \centering
  \includegraphics[width=\linewidth]{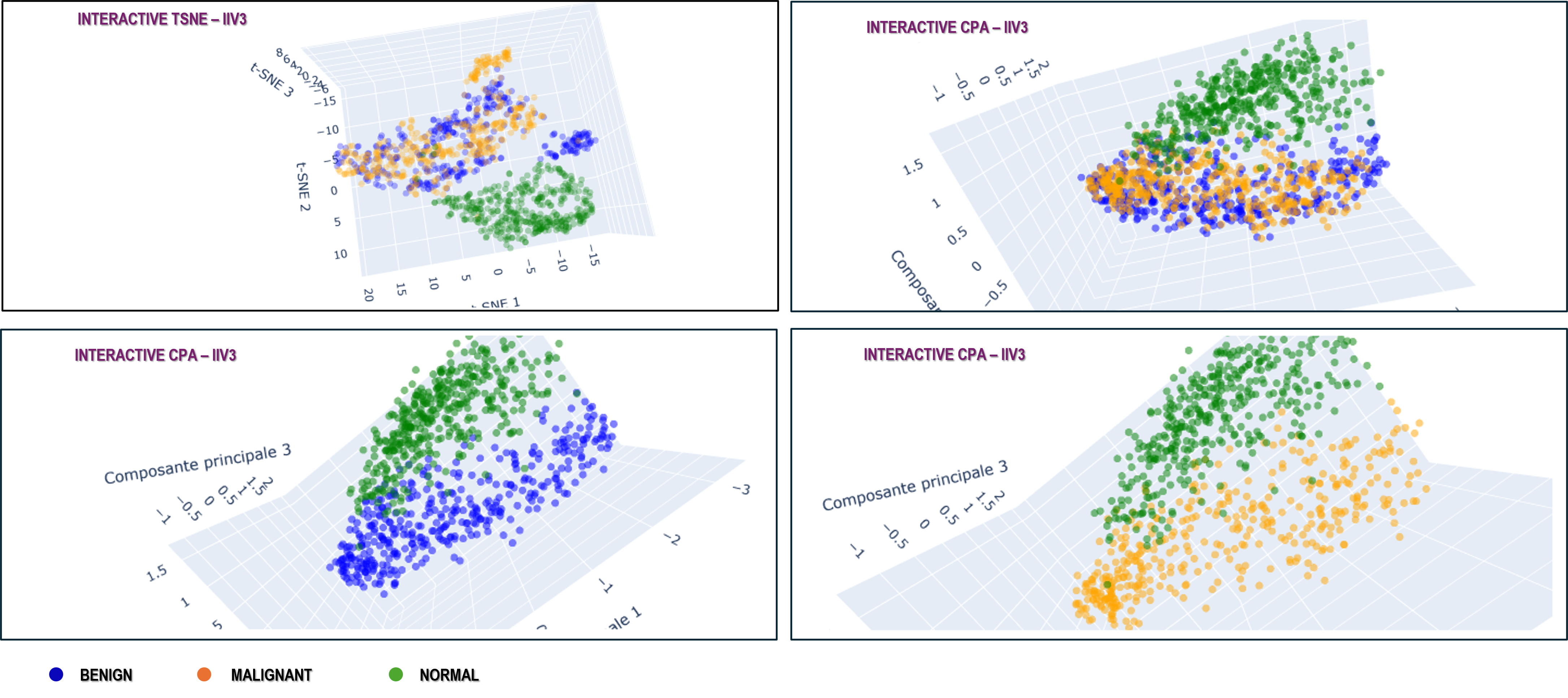}
  \caption{Latent Space Projections of Improved InceptionV3 Using PCA and t‑SNE. Compact clusters with reduced intra‑class variability highlight the model’s stability and robustness, even though benign and malignant classes still exhibit partial overlap.}
  \label{fig13_separability_iiv3}
\end{figure}

\textbf{Improved InceptionV3.}  
InceptionV3 encoded features in a denser representation, with PCA curves (notably from \texttt{dense\_1}) showing rapid variance capture. The 3D PCA and t‑SNE projections (Figure~\ref{fig13_separability_iiv3}) revealed compact clusters with reduced intra‑class variability. However, the separation between benign and malignant was not dramatically superior to ICNT; overlaps remained visible. The strength of InceptionV3 lies more in the compactness and stability of its latent space than in visibly sharper boundaries. This dense structuring likely contributes to its robustness and generalization capacity, even if visual separability is not radically clearer than ICNT.  

\textbf{Synthesis}  
Latent space analyses reveal complementary strengths across models. The baseline CNN provides a simple but less discriminative representation, with evident overlaps between critical classes. ICNT improves separability, offering clearer boundaries and more interpretable latent clusters. InceptionV3, while not visually more distinct than ICNT, achieves superior compactness and internal organization, which aligns with its strong quantitative performance and robustness. These findings underscore that latent space evaluation is not only about visual separability but also about structural compactness, both of which are crucial for reliable clinical deployment.
\subsection{Model Explainability and Visual Attention (Grad-CAM)}
To complement quantitative and latent space analyses, local interpretability was assessed using Gradient‑weighted Class Activation Mapping (Grad‑CAM). This technique highlights the regions of mammograms that most influenced the models’ predictions, thereby providing insight into the decision‑making process and addressing the “black box” concern often raised in medical AI.  

\begin{figure}[H]
  \centering
  \includegraphics[width=\linewidth]{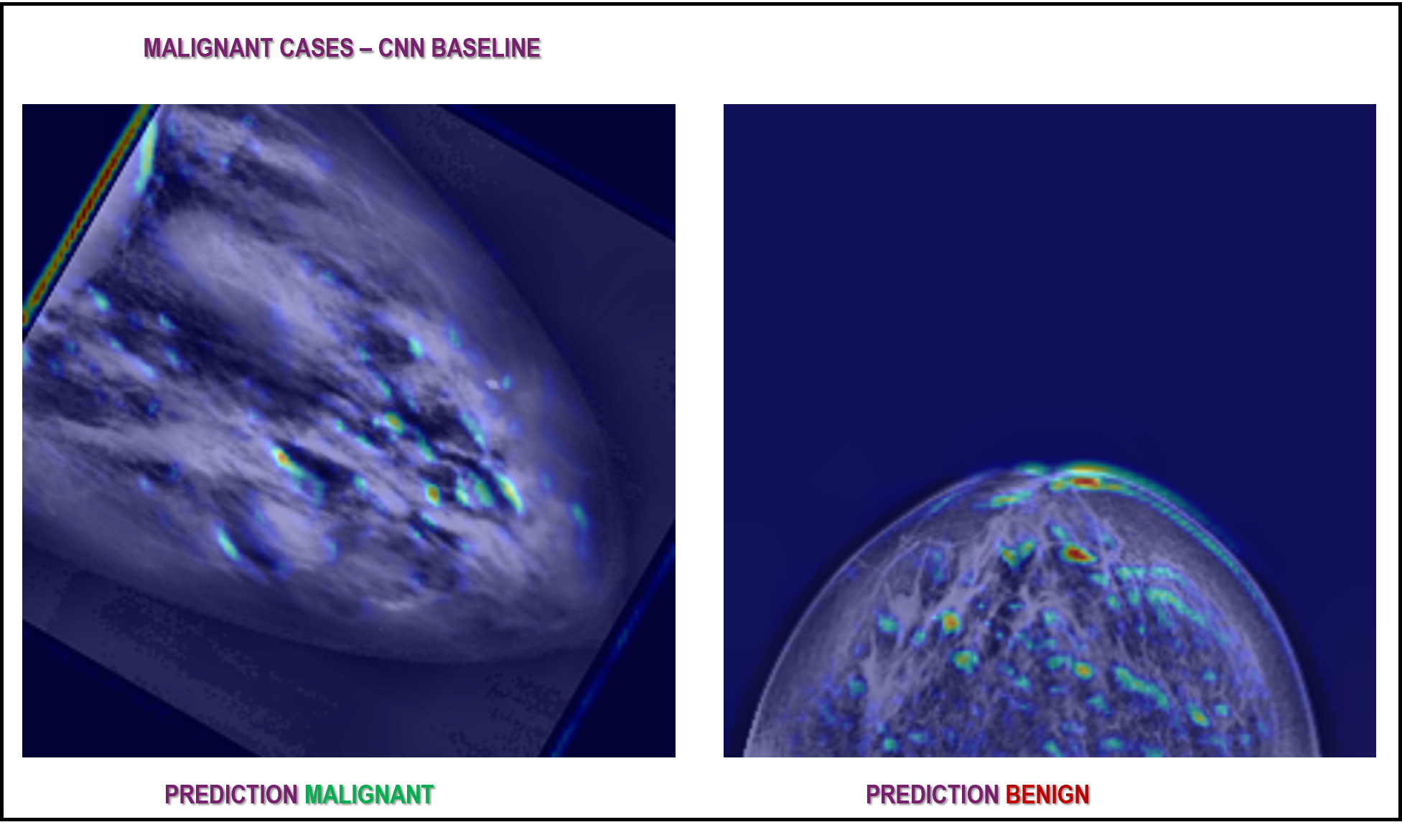}
  \caption{Grad‑CAM Visualizations for Malignant Cases Using the Baseline CNN. Correctly classified malignant cases show narrow and inconsistent activations, while misclassified cases reveal misplaced or insufficient attention, underscoring the limited representational capacity of the baseline CNN.}
  \label{fig14_cnn_gradcam}
\end{figure}

\textbf{Baseline CNN.}  
Grad‑CAM visualizations for the baseline CNN (Figure~\ref{fig14_cnn_gradcam}) revealed limited and highly localized activations. In correctly classified malignant cases, heatmaps highlighted small regions overlapping with suspicious tissue, but the coverage was narrow and inconsistent. In critical errors, such as malignant cases misclassified as benign, activations were misplaced or insufficient to discriminate the class. This limitation reflects the restricted representational capacity of the baseline CNN.  

\begin{figure}[H]
  \centering
  \includegraphics[width=\linewidth]{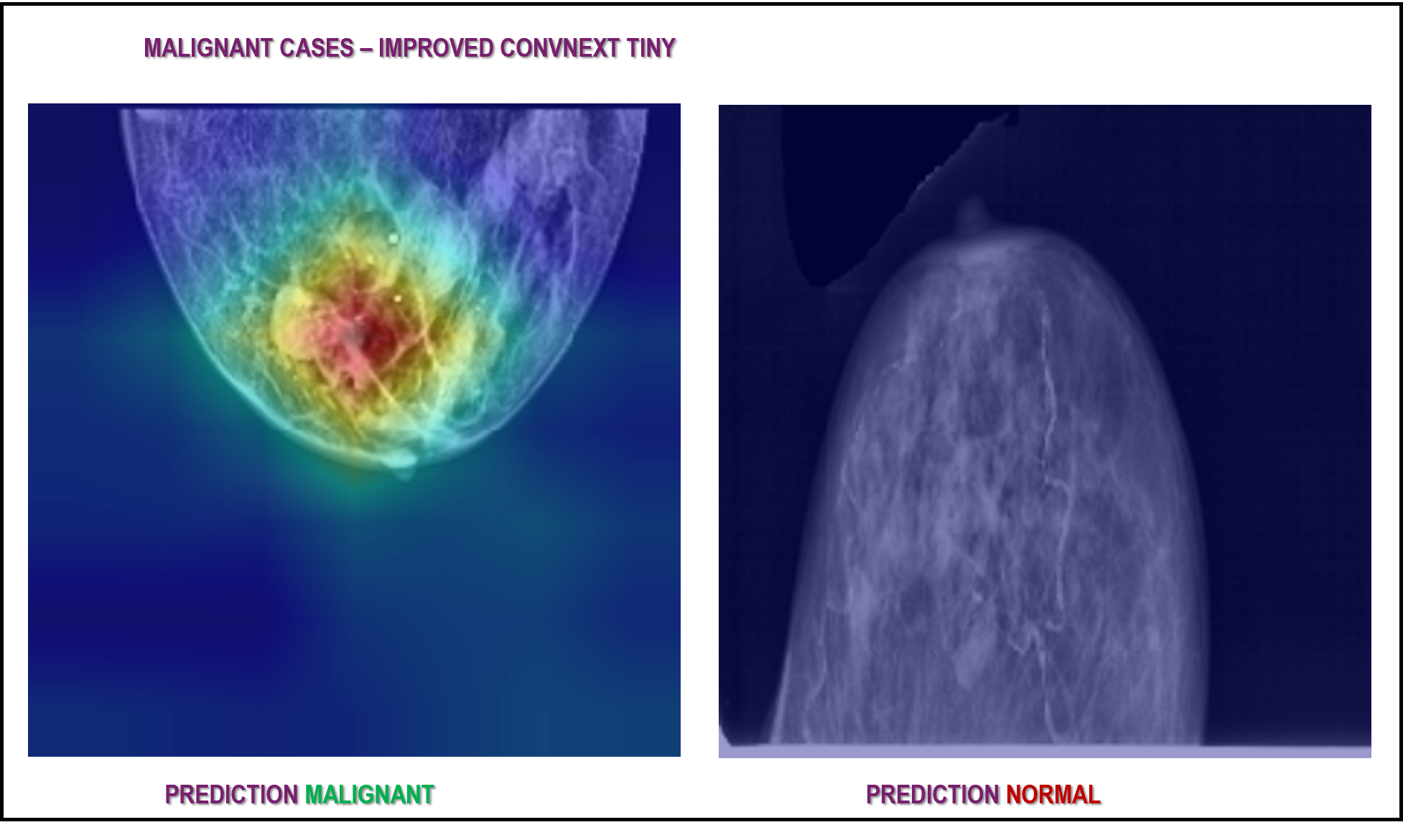}
  \caption{Grad‑CAM Visualizations for Malignant Cases Using Improved ConvNeXt Tiny. Correct predictions highlight regions adjacent to visible lesions, while critical errors reveal complete absence of relevant activations, underscoring the risk of false negatives in clinical practice.}
  \label{fig15_icnt_gradcam}
\end{figure}

\textbf{Improved ConvNeXt Tiny (ICNT).}  
ICNT produced more informative visualizations in correct predictions, with heatmaps highlighting regions adjacent to visible lesions (Figure~\ref{fig15_icnt_gradcam}). However, in critical errors, such as malignant cases misclassified as normal with high confidence, Grad‑CAM maps showed no relevant activation at all. This “blindness” explains the failure and underscores the risk of false negatives in clinical practice.  

\begin{figure}[H]
  \centering
  \includegraphics[width=\linewidth]{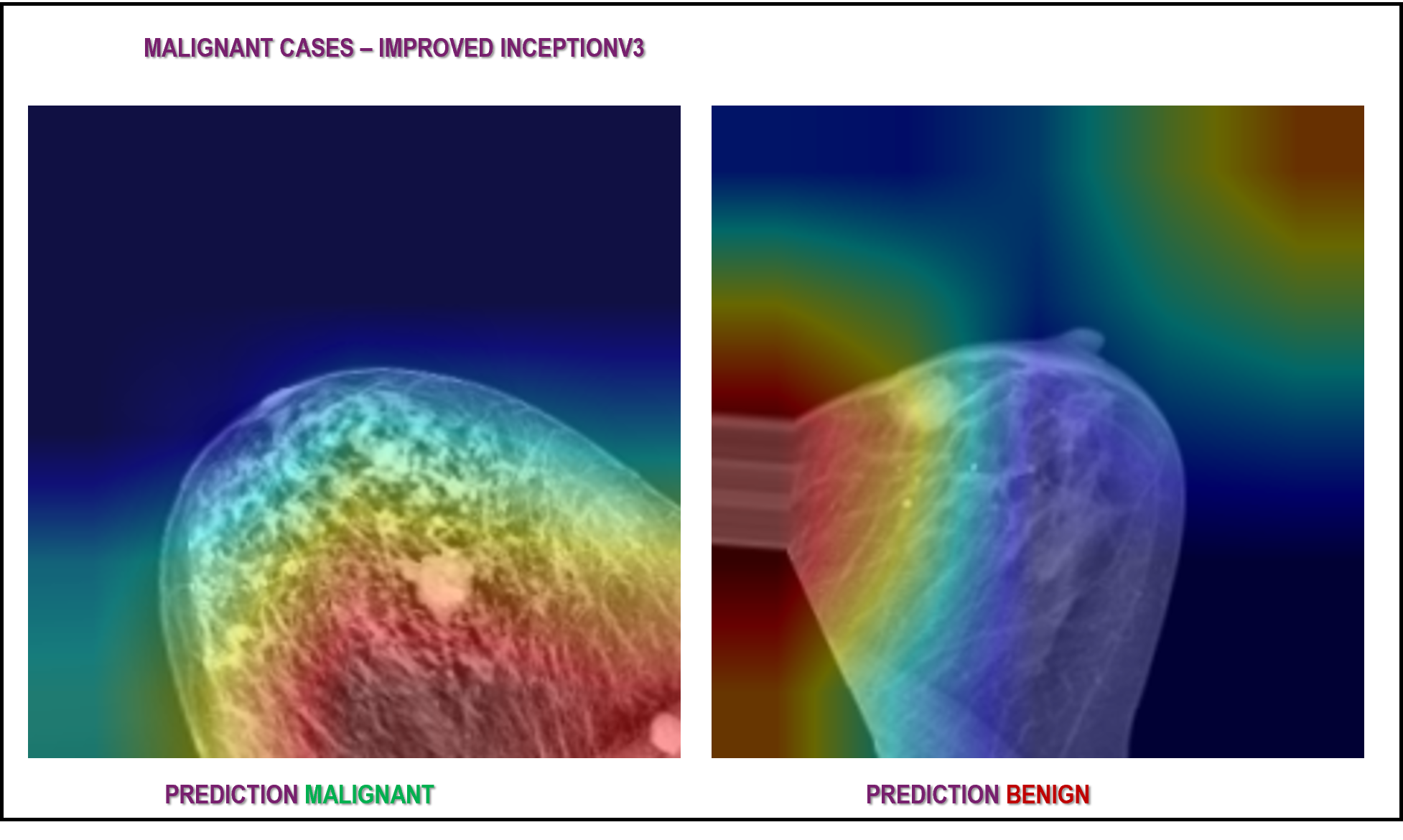}
  \caption{Grad‑CAM Visualizations for Malignant Cases Using Improved InceptionV3. Correct predictions show activations superimposed directly on tumor regions, while misclassifications reveal insufficient but present activations, highlighting the model’s relative reliability in focusing on clinically meaningful areas.}
  \label{fig16_inceptionv3_gradcam}
\end{figure}

\textbf{Improved InceptionV3.}  
InceptionV3 demonstrated the most robust and clinically coherent attention patterns. In correctly classified malignant cases, Grad‑CAM heatmaps (Figure~\ref{fig16_inceptionv3_gradcam}) were superimposed directly on tumor regions, confirming that the model focused on relevant pathology. In misclassifications, such as malignant cases predicted as benign, activations were still present but insufficient to discriminate the class. This shows that attention does not guarantee understanding, but InceptionV3 remains more reliable than the other models in aligning activations with clinically meaningful regions.  

\textbf{Synthesis and Limitations}  
These examples illustrate a fundamental limitation of Grad‑CAM: it shows where the model concentrates its attention, but not why it makes a decision. Even when models misclassify, they may activate relevant areas, which demonstrates that attention alone does not equate to comprehension. This is a critical limitation in medical contexts, where transparency of reasoning is essential for trust and adoption. Grad‑CAM provides valuable visual cues, but it cannot fully explain the decision process, and must be complemented by global interpretability methods.

\section{Conclusion and Perspectives}

This proof of concept provided an opportunity to explore the state of the art in deep learning applied to mammography analysis. The implementation of enhancement modules such as GAGM and SEVector proved promising for improving feature discriminability and reducing false negatives, while the Feature Smoothing Loss (FSL) showed limited relevance in our context. This contrasts with Xia et al.~\cite{xia2025convnexttiny}, who demonstrated its effectiveness in Alzheimer’s MRI under CPU‑friendly conditions, suggesting that its utility may depend on task‑specific architectural and computational assumptions. Our results therefore both validate and extend their framework, confirming the transposability of GAGM and SEVector while critically reassessing the role of FSL.  

Beyond experimental validation, clinical deployment requires careful consideration of use cases and operational constraints. For initial diagnosis, where minimizing false negatives is paramount, Improved InceptionV3 remains the most suitable choice despite its computational cost. For longitudinal follow‑up or mass screening, however, lighter models such as the baseline CNN or ICNT may be preferable, as they offer faster inference and easier maintenance even if their accuracy is lower. These scenarios illustrate that model selection must balance diagnostic accuracy with feasibility in real healthcare environments.  

Explainability remains a major challenge. Grad‑CAM provided valuable local insights, but its effectiveness varied across models and cannot fully explain decision processes. Classical approaches such as LIME, SHAP, occlusion maps, or concept activation vectors have been widely adopted, yet they also face well‑documented limitations in terms of stability, faithfulness, and clinical interpretability. Future work should therefore move beyond these traditional tools and explore more recent perspectives, including prototype‑based explanations, counterfactual reasoning, and concept bottleneck models~\cite{salih2023shaplime,wang2024interpretabilityframework}. Such approaches hold greater promise for delivering clinically meaningful and trustworthy insights into model reasoning, aligning interpretability with the demands of real healthcare environments.  

Another limitation concerns image heterogeneity. Models trained on homogenized, preprocessed mammograms may behave unpredictably on data of different quality. Strengthening robustness to variability is essential for clinical generalization, and external validation on multi‑center datasets will be necessary to confirm reliability in diverse conditions. Finally, our attempt to implement FSL highlighted the difficulty of transposing certain methodological bricks from the literature without adaptation. Future iterations may explore simplified variants or combinations with other regularization strategies to assess their true added value.  

In summary, Improved InceptionV3 emerged as the optimal model in this experimental setting, but the final choice of architecture must balance diagnostic accuracy with operational feasibility. Future research should aim to design models that are not only performant but also explainable, lightweight, and adaptable to diverse healthcare infrastructures, ensuring equitable access to advanced diagnostic tools worldwide.
\section*{Acknowledgements}
The author would like to express sincere gratitude to Koen Bertels and Imen Turki for their valuable methodological guidance and constructive feedback throughout this work. Special thanks are also extended to Catalina Chircu for her evaluation of the proof‑of‑concept project on mammography classification.  

This study further benefited from the use of several technical platforms, including Google Colab, Kaggle, JupyterHub, HuggingFace, and Streamlit, which provided essential computational resources and tools for experimentation and visualization.

\section*{Availability of the Dashboard}
As part of this proof of concept, an interactive dashboard has been deployed on Streamlit Cloud and is accessible at \url{https://pocmedicalimagesclassification.streamlit.app/}. 
This deployment illustrates the feasibility of web‑based exploration of model outputs. 
Due to resource limitations inherent to Streamlit Community Cloud, availability may be intermittent. 
Future iterations will consider more robust hosting solutions to ensure stability and scalability.
\bibliographystyle{IEEEtran}
\bibliography{references}

@article{xia2025convnexttiny,
  author    = {Xia, Jingsong and Yin, Yue and Li, Xiuhan},
  title     = {An Efficient Medical Image Classification Method Based on a Lightweight Improved ConvNeXt-Tiny Architecture},
  journal   = {arXiv preprint arXiv:2508.11532 [cs.CV]},
  year      = {2025},
  month     = aug,
  doi       = {10.48550/arXiv.2508.11532},
  url       = {https://arxiv.org/abs/2508.11532}
}

@misc{venegas2023kaggle,
  author       = {Venegas, Emilio},
  title        = {Mammography dataset from INbreast, MIAS, and DDSM},
  howpublished = {Kaggle},
  year         = {2023},
  url          = {https://www.kaggle.com/datasets/emiliovenegas1/mammography-dataset-from-inbreast-mias-and-ddsm}
}

@misc{iarc2022gco,
  author       = {{International Agency for Research on Cancer (IARC)}},
  title        = {Global Cancer Observatory: Cancer Today},
  howpublished = {World Health Organization, Lyon, France},
  year         = {2022},
  url          = {https://gco.iarc.fr}
}

@article{kim2025globalpatterns,
  author    = {Kim, J. and Harper, A. and McCormack, V. and Sung, H. and Houssami, N. and Morgan, E. and others},
  title     = {Global patterns and trends in breast cancer incidence and mortality across 185 countries},
  journal   = {Nature Medicine},
  year      = {2025},
  note      = {Published online Feb. 24, 2025},
  doi       = {10.1038/s41591-025-03502-3}
}

@article{shi2025screeningvalue,
  author    = {Shi, J. and Li, J. and Gao, Y. and Chen, W. and Zhao, L. and Li, N. and Tian, J. and Li, Z.},
  title     = {The screening value of mammography for breast cancer: an overview of 28 systematic reviews with evidence mapping},
  journal   = {Journal of Cancer Research and Clinical Oncology},
  volume    = {151},
  pages     = {102},
  year      = {2025},
  month     = mar,
  doi       = {10.1007/s00432-025-06122-z},
  url       = {https://link.springer.com/article/10.1007/s00432-025-06122-z}
}

@article{henderson2024jama,
  author    = {Henderson, J. T. and Webber, E. M. and Weyrich, M. S. and others},
  title     = {Screening for Breast Cancer: Evidence Report and Systematic Review for the US Preventive Services Task Force},
  journal   = {JAMA},
  volume    = {331},
  number    = {22},
  pages     = {1931--1946},
  year      = {2024},
  month     = apr,
  url       = {https://jamanetwork.com/journals/jama/fullarticle/2818284}
}

@article{hassan2022cadreview,
  author    = {Hassan, N. M. and Hamad, S. and Mahar, K.},
  title     = {Mammogram breast cancer CAD systems for mass detection and classification: a review},
  journal   = {Multimedia Tools and Applications},
  volume    = {81},
  pages     = {20043--20075},
  year      = {2022}
}

@article{hussain2025optimizeddl,
  author    = {Hussain, S. I. and Toscano, E.},
  title     = {Optimized Deep Learning for Mammography: Augmentation and Tailored Architectures},
  journal   = {Information},
  volume    = {16},
  number    = {5},
  pages     = {359},
  year      = {2025}
}

@article{sharma2025comparative,
  author    = {Sharma, S. and Singh, Y. and Choudhury, T.},
  title     = {Advanced deep learning architectures for enhanced mammography classification: a comparative study of CNNs and ViT},
  journal   = {Discover Artificial Intelligence},
  volume    = {5},
  number    = {187},
  year      = {2025}
}

@article{alantari2024cad,
  author    = {Al-Antari, M. A. and Al-Masni, M. A. and Kim, T.-S.},
  title     = {Deep learning computer-aided diagnosis for breast cancer detection from mammography images},
  journal   = {Frontiers in Oncology},
  volume    = {14},
  pages     = {1281922},
  year      = {2024}
}

@article{selvaraju2016gradcam,
  author    = {Selvaraju, R. R. and Cogswell, M. and Das, A. and Vedantam, R. and Parikh, D. and Batra, D.},
  title     = {Grad-CAM: Visual Explanations from Deep Networks via Gradient-Based Localization},
  journal   = {arXiv preprint arXiv:1610.02391 [cs.CV]},
  year      = {2016},
  month     = oct,
  url       = {https://arxiv.org/abs/1610.02391}
}

@article{li2025resgdanet,
  author    = {Li, S. and Huang, J.},
  title     = {ResGDANet: An Efficient Residual Group Attention Neural Network for Medical Image Classification},
  journal   = {Applied Sciences},
  volume    = {15},
  number    = {5},
  pages     = {2693},
  year      = {2025},
  doi       = {10.3390/app15052693},
  url       = {https://www.mdpi.com/2076-3417/15/5/2693}
}

@article{hu2017senet,
  author    = {Hu, J. and Shen, L. and Sun, G.},
  title     = {Squeeze-and-Excitation Networks},
  journal   = {arXiv preprint arXiv:1709.01507 [cs.CV]},
  year      = {2017},
  month     = sep,
  url       = {https://arxiv.org/abs/1709.01507}
}

@article{wen2016discriminative,
  author    = {Wen, Y. and Zhang, K. and Li, Z. and Qiao, Y.},
  title     = {A Discriminative Feature Learning Approach for Deep Face Recognition},
  journal   = {arXiv preprint arXiv:1604.01345 [cs.CV]},
  year      = {2016},
  month     = apr,
  url       = {https://arxiv.org/abs/1604.01345}
}

@article{roy2023mednext,
  author    = {Roy, S. and Koehler, G. and Ulrich, C. and others},
  title     = {MedNeXt: Transformer-driven Scaling of ConvNets for Medical Image Segmentation},
  journal   = {arXiv preprint arXiv:2303.09975 [cs.CV]},
  year      = {2023},
  month     = mar,
  url       = {https://arxiv.org/abs/2303.09975}
}

@article{hatamizadeh2022unetrpp,
  author    = {Hatamizadeh, A. and Tang, Y. and Nath, V. and others},
  title     = {UNETR++: Delving into Efficient and Accurate 3D Medical Image Segmentation},
  journal   = {arXiv preprint arXiv:2209.04462 [cs.CV]},
  year      = {2022},
  month     = sep,
  url       = {https://arxiv.org/abs/2209.04462}
}

@article{alam2024smote,
  author    = {Alam, M. Z. and Roy, T. and Kawsar, H. M. N. and Rimi, I.},
  title     = {Enhancing Transfer Learning for Medical Image Classification with SMOTE: A Comparative Study},
  journal   = {arXiv preprint arXiv:2412.20235 [cs.CV]},
  year      = {2024},
  month     = dec,
  url       = {https://arxiv.org/abs/2412.20235}
}

@article{deng2014imagenet,
  author    = {Deng, J. and Dong, W. and Socher, R. and Li, L.-J. and Li, K. and Fei-Fei, L.},
  title     = {ImageNet: A large-scale hierarchical image database},
  journal   = {arXiv preprint arXiv:1409.0575 [cs.CV]},
  year      = {2014},
  month     = sep,
  url       = {https://arxiv.org/abs/1409.0575}
}

@article{srivastava2012dropout,
  author    = {Srivastava, N. and Hinton, G. and Krizhevsky, A. and Sutskever, I. and Salakhutdinov, R.},
  title     = {Dropout: A Simple Way to Prevent Neural Networks from Overfitting},
  journal   = {arXiv preprint arXiv:1207.0580 [cs.NE]},
  year      = {2012},
  month     = jul,
  url       = {https://arxiv.org/abs/1207.0580}
}

@article{ioffe2015batchnorm,
  author    = {Ioffe, S. and Szegedy, C.},
  title     = {Batch Normalization: Accelerating Deep Network Training by Reducing Internal Covariate Shift},
  journal   = {arXiv preprint arXiv:1502.03167 [cs.LG]},
  year      = {2015},
  month     = feb,
  url       = {https://arxiv.org/abs/1502.03167}
}

@article{salih2023shaplime,
  author    = {Salih, A. and Khan, M. and Ahmad, R.},
  title     = {A Perspective on Explainable AI Methods: SHAP and LIME},
  journal   = {arXiv preprint arXiv:2305.02012 [cs.LG]},
  year      = {2023},
  month     = may,
  url       = {https://arxiv.org/abs/2305.02012}
}

@article{wang2024interpretabilityframework,
  author    = {Wang, Y. and Li, J. and Zhang, H.},
  title     = {A Framework for Interpretability in Machine Learning for Medical Imaging},
  journal   = {arXiv preprint arXiv:2310.01685 [cs.CV]},
  year      = {2024},
  month     = oct,
  url       = {https://arxiv.org/abs/2310.01685}
}

\end{document}